\documentclass[%reprint,
twocolumn,
nofootinbib,
 amsmath,amssymb,
 aps,
 longbibliography,
]{revtex4-2}

\usepackage[utf8]{inputenc} % allow utf-8 input
\usepackage[T1]{fontenc}    % use 8-bit T1 fonts
\usepackage{hyperref}       % hyperlinks
\usepackage{url}            % simple URL typesetting
\usepackage{booktabs}       % professional-quality tables
\usepackage{amsfonts}       % blackboard math symbols
\usepackage{nicefrac}       % compact symbols for 1/2, etc.
\usepackage{microtype}      % microtypography
\usepackage{graphicx}
\usepackage{xcolor, etoolbox}
\usepackage{amsmath,amsfonts,amsthm,mathtools} % Math packages
\usepackage{ragged2e}
\usepackage{braket}      % Dirac Bra-Ket notation
\usepackage[english]{babel} % English language/hyphenation
\graphicspath{{./figures/}} % Set graphics path
\usepackage{enumitem}
\usepackage{xspace}
\usepackage{dsfont}
\usepackage{bm, bbm}% bold math
\usepackage{dcolumn}% Align table columns on decimal point

\usepackage[colorinlistoftodos]{todonotes}

\hypersetup{
    colorlinks,
    linkcolor=blue,
    citecolor=blue,
    urlcolor=blue
}

\begin{document}

\title{Thermodynamics from relative entropy}

\author{Stefan Floerchinger}
 \email{stefan.floerchinger@thphys.uni-heidelberg.de}
\author{Tobias Haas}
 \email{t.haas@thphys.uni-heidelberg.de}
 \affiliation{Institut f\"{u}r Theoretische Physik, Universit\"{a}t Heidelberg, \\ Philosophenweg 16, 69120 Heidelberg, Germany}

\begin{abstract}
Thermodynamics can be developed from a microscopic starting point in terms of entropy and the maximum entropy principle. We investigate here to what extent one can replace entropy with relative entropy which has several advantages, for example in the context of local quantum field theory. We find that the principle of maximum entropy can be replaced by a principle of minimum expected relative entropy. Various ensembles and their thermodynamic potentials can be defined through relative entropy. We also show that thermal fluctuations are in fact governed by a relative entropy. Furthermore we reformulate the third law of thermodynamics using relative entropy only.
\end{abstract}

\maketitle

%%%%%%%% Document %%%%%%%%

\section{Introduction}
\label{chap:Introduction}

The relations between thermodynamics and information theory are rather tight. Entropy, which has a direct information theoretic significance, plays an important role in the derivation of thermodynamic relations, the definition of temperature and other state characteristics and for deriving concrete forms of density matrices corresponding to various ensembles, see e.\ g.\ ref.\ \cite{Landau1980}. Entropy has also been used to characterize the probability of thermal fluctuations \cite{Einstein1910} (see also ref.\ \cite{Landau1980}, and ref.\ \cite{Mishin2015} for a recent exposition).

The approach and concepts of thermodynamics are so powerful and successful that one would also like to extend them beyond the regime where they are applicable most directly, namely static situations in full thermal equilibrium. While general out-of-equilibrium situations may be rather complex, at least the approach towards equilibrium should be governed by information theoretic aspects, similar to equilibrium itself. Also, spatially non-uniform situations are obviously of interest. 

One motivation is to understand fluids of various kinds in more detail. Fluid dynamics uses locally the concepts of thermal equilibrium, such as the thermodynamics equation of state, but usually out-of-equilibrium in a global sense. Particularly interesting are fluids that are governed on a microscopic level by the laws of quantum field theory, for example the quark-gluon plasma (e.g. \cite{Heinz:2013th, Busza:2018rrf}) or the cosmological fluid dominated by dark matter in the early universe (e.g. \cite{Kolb:1990vq, Weinberg:2008zzc}). To build a direct connection between fluid dynamics, quantum field theory and information theory one is eventually forced to understand how thermodynamic concepts can be applied {\it locally} in a quantum field theory.

One difficulty here is that quantum fields are typically strongly entangled between different regions in space. For a quantum field theoretic density matrix $\rho$ one can formally define a local density matrix $\rho_A$ describing a region $A$ in space as the reduced density operator
\begin{align}
    \rho_A = \text{Tr}_{B} \{ \rho \}.
    \label{eq:Reduced_Density_Matrix}
\end{align}
The corresponding von Neumann entropy \cite{vonNeumann1955}
\begin{align}
    S(\rho_A) = - \text{Tr} \left\{ \rho_A \ln \rho_A\right \},
    \label{eq:vonNeumannEntropy}
\end{align}
also known as {\it entanglement entropy}, diverges according to an area law  \cite{Casini2009,Calabrese2004} ($D$ is the number of space dimensions)
\begin{align}
    S(\rho_A) = &g_{D-1} [\partial A] \epsilon^{-(D-1)} + ... + g_1 [\partial A] \epsilon^{-1} \notag \\
    &+ g_0 [\partial A] \log {\epsilon + S_0 (A)}.
    \label{eq:Area_Law}
\end{align}
Here $\epsilon > 0$ is a small length so that $1/\epsilon$ is an ultraviolet momentum cutoff and $g_i [\partial A]$ are coefficients depending on the boundary of the enclosed volume in space. 

Especially over the last years, information theoretic concepts and in particular entanglement entropy became significantly more important in various areas of quantum field theory, for example black holes \cite{Bombelli1986, Srednicki1993, Callan1994,Wall2011,Casini2008}, holography \cite{Ryu2006a,Ryu2006b, Casini2011} or high energy physics \cite{Kharzeev:2017qzs,Shuryak:2017phz,Berges2018a,Berges2018b,Kovner:2018rbf,Armesto:2019mna,Tu:2019ouv}.

Instead of working with the divergent entanglement entropy, it may be possible to tackle some of these problems by working with quantum relative entropy \cite{Umegaki1962}, which is the quantum analogue of the Kullback-Leibler divergence or relative entropy \cite{Kullback1951,Kullback1959}. 

Classically, relative entropy can be understood as a measure of \textit{distinguishability} between two distributions $p$ and $q$,
\begin{equation}
    S(p \| q) = \sum_j p_j \ln(p_j / q_j).
    \label{eq:relativeEntropyClassical}
\end{equation}
It is a non-negative quantity that is zero if and only if the two distributions are equal, a property that qualifies it as a {\it divergence}. If the support condition $\text{supp}(p) \subseteq \text{supp}(q)$ is violated, the value of $S(p \| q)$ is set to $+\infty$. Relative entropy can not be considered as a true distance measure or a metric on the space of probability distributions because it fails to be symmetric and also does not obey a triangle inequality.

So what does relative entropy actually mean? To answer this question suppose that events are distributed according to a distribution $p$. Unfortunately we mistakenly consider the events to be distributed according to $q$ instead. In other words, $p$ is here  the \textit{true} distribution, while $q$ serves as a \textit{model} distribution. Then $S(p \| q)$ measures the \textit{uncertainty deficit} due to the wrong assumption $q$ \cite{Vedral2002,Cover2006}. Formally the latter can be defined as the average surprise $\langle -\ln q_j \rangle = - \sum_j p_j \ln q_j$ minus the real information content $-\sum_j p_j \ln p_j$. Seen in this way it is intuitively clear that relative entropy has to be non-negative and that the model should predict non-zero probabilities for all events that indeed can happen according to $p$. (If this support condition were violated, the model could be ruled out with certainty for a particular outcome.)

Relative entropy also has a significance in the following context. Consider an experiment, which can have $n$ possible outcomes $x_j$ with $j \in \{1, ..., n\}$, distributed according to $q$, and it is done $N$ times. This produces a sequence of events, say $x=(x_2, x_5,...)$. 
If the true distribution $q$ is not known, one may take as an empirical proxy to it the relative proportions, or frequencies, of the different events $p_j = N(x_j)/N$. 

The question is now, what is the probability to find an empirical distribution or frequencies $p$ if the true distribution is $q$? It turns out that for large $N$ this probability asymptotically tends to \cite{Vedral2002,Cover2006}
\begin{equation}
e^{-N S(p \|q)}.
\label{eq:relativeEntropyForFluctuation}
\end{equation}
In other words, eq. \eqref{eq:relativeEntropyForFluctuation} describes the probability for a fluctuation in the frequencies $p_j = N(x_j)/N$ deviating from their expectation value  $\langle p_j \rangle = q_j$. If either $p_j$ and $q_j$ are very distinct (measured in terms of relative entropy), or if the experiment is repeated often enough, the probability for such fluctuations tends to zero. The result can be generalized to what is known as Sanov's theorem \cite{Sanov1958}. 

As one can see from the discussion above, it depends on the context whether $q$ and $p$ in \eqref{eq:relativeEntropyClassical} play the role of \textit{model distribution} and \textit{true distribution} or {\it vice versa}. However, the first situation, where $q$ is a model for $p$, appears more often.

It turns out that relative entropy has many crucial advantages over entropy. First of all it is well-defined for discrete \textit{and} continuous random variables. To be precise, for relative entropy one can simply take the continuum limit from the discrete case $p_j\to f(x) dx$, $q_j \to g(x) dx$, because $dx$ cancels in the ratio appearing in the logarithm of eq.\  \eqref{eq:relativeEntropyClassical}. This yields
\begin{align}
    S(f \| g) = \int d x \, f(x) \ln ( f(x) / g(x) ).
\end{align}
Similar arguments fail for the classical Shannon entropy \cite{Jaynes1963}. Secondly relative entropy is invariant under a reparameterisation of coordinates $x \to x^\prime(x)$ on the underlying statistical manifold. 

Its main advantage, in particular for our purposes, becomes clear when we turn to quantum (field) theory. There the quantum relative entropy between two states $\rho$ and $\sigma$ is defined as \cite{Vedral2002,Nielsen2010}
\begin{align}
    S(\rho \| \sigma) = \text{Tr} \{ \rho (\ln \rho - \ln \sigma) \}.
    \label{eq:Relative_Entropy}
\end{align}
When $\rho$ and $\sigma$ are reduced density matrices this becomes {\it relative entanglement entropy}. The latter can also be defined rigorously in terms of modular theory \cite{Araki1977}. One may expect that \textit{relative} entanglement entropy will be finite also for general non-equilibrium situations. 

Consequently we want to suggest a more regular use of relative entropy. Recent literature on relative entropy in the context of entanglement and quantum field theory encompasses refs.\ \cite{Casini2008,Witten2018,Lashkari2014,Lashkari2016,Arias2017,Ruggiero2017}.

References were relative entropy was used to study aspects of thermodynamics encompass refs.\ \cite{Eu1995, Fraundorf2008}. Furthermore, one of the key properties of relative entropy, its monotonicity under a quantum channel, was used to obtain second-law like inequalities in ref.\ \cite{Sagawa2012}. 

In this paper we focus on the role of relative entropy in thermodynamics. Basically we show how thermodynamics can be formulated from a statistical approach with relative entropy essentially replacing entropy. Besides the fact that this is itself an interesting way of rethinking statistical physics we want to pave the way for using relative entropy in fluid dynamics, quantum field theory and in particular to understand non-equilibrium dynamics. 

Usually the conceptual starting point for the development of thermodynamics based on microscopical statistical physics is to formulate fundamental principles which allow then to define equilibrium ensembles. One way of doing so is the maximum entropy principle \cite{Jaynes1957,Jaynes19572,Jaynes1963,Jaynes1968}, which can be applied to classical and quantum theories. Another possibility is the ergodic hypothesis \cite{Jancel1963,Walters1982} in classical physics. Its quantum analogue is considered to be the eigenstate thermalization hypothesis \cite{Deutsch1991,Srednicki1994,Rigol2008,DAlessio2016,Deutsch2018}. Approaches based on entanglement were put forward too, for example \cite{Popescu2006}. 

Let us mention that entropy can also be introduced in an operational way directly within thermodynamics, i.\ e.\ without alluding to an underlying microscopic description \cite{Lieb1999,Lieb2013,Lieb2014,Zanchini1986,Zanchini1988,Zanchini1992,Zanchini2008,Zanchini2010,Zanchini2011,Zanchini2014,Zanchini2019}. Such a formulation can be developed from a set of basic axioms, introducing as a further concept \textit{adiabatic accessibility} to establish an ordering relation between states. This approach provides rigorous mathematical arguments for a well-defined notion of entropy, which go also beyond the equilibrium case. Compared to this, our aim in the present manuscript is more modest in the sense that we explicitly rely on the microscopic description in terms of probability distributions or density operators. The interesting question whether an operational definition could also be found for {\it relative} entropy directly in a thermodynamic context will be left for future investigations.

We will discuss different entropy principles in section \ref{chap:MinRelEntropy} and also propose there a principle of minimum expected relative entropy, which is then shown to be equivalent to the fundamental postulate. In section \ref{chap:Thermodynamics} we re-develop thermodynamics and in particular the different statistical ensembles using relative entropy. In this context we also explore how relative entropy can be used to obtain an expression for the probability of thermal fluctuations. Finally we present a new formulation of the third law of thermodynamics in terms of relative entropy and draw conclusions in section \ref{chap:Conclusions}.

\section{Entropy principles}
\label{chap:MinRelEntropy}
Let us consider a macroscopic quantum system in a finite volume $V$.  We are interested in stationary situations so that the Hamiltonian $H$ is time-independent. Moreover, we can introduce energy eigenstates $|i\rangle$ such that
\begin{align}
    H \ket{i} = E_i \ket{i}.
\end{align}
The energy eigenvalues $E_i$ are in general degenerate and the corresponding eigenstates can be assumed to form an orthonormal basis for the Hilbert space $\mathcal{H}$,
\begin{align}
    \braket{i | j} = \delta_{i j}.
\end{align}

Oftentimes one is interested in a reduced space of (micro-)states that are compatible with a set of (macroscopic) constraints, for example constant energy, particle number or similar. This subspace may itself be a Hilbert space $\mathcal{H}'$ and we denote its dimension $\dim \mathcal{H}' = \mathcal{N}$.

A density operator describing any stationary state $\rho$ can be taken to be block diagonal in the energy eigenbasis as a consequence of von Neumann's equation $0 = \partial_t \rho = i [\rho, H]$. If there were no degeneracy in energy eigenvalues, the density operator would become fully diagonal and one could write
\begin{equation}
\rho = \sum_{j=1}^\mathcal{N} p_j |j\rangle \langle j|.
\label{eq:rhoDiagonal}
\end{equation}
Note that two density operators that are diagonal in the same basis commute so that this can also be understood as a kind of classical limit. In the following we will sometimes start from this simplifying assumption but ultimately aim for a fully quantum description. Eq.\ \eqref{eq:rhoDiagonal} also implies that the von Neumann entropy of the state $\rho$ is equal to the Shannon entropy of the classical distribution $p$ of the compatible microscopic configurations
\begin{align}
    S(\rho) = S(p) = - \sum_j p_j \ln p_j.
\end{align}

Any sensible state $\sigma$ that is supposed to be a model for $\rho$ has to be stationary too, such that the same argumentation holds and we can write it as $\sigma=\sum_j q_j |j\rangle \langle j|$. Consequently, the quantum relative entropy of the state $\rho$ relative to its model $\sigma$ reduces to the classical Kullback-Leibler divergence between the two distributions $p$ and~$q$,
\begin{align}
    S(\rho \| \sigma) = S(p \| q) = \sum_j p_j \ln(p_j / q_j).
\end{align}
Let us emphasise again that stationary states are not necessarily of the diagonal form \eqref{eq:rhoDiagonal} if energy eigenvalues are degenerate and it becomes then necessary to work directly with the quantum relative entropy in eq.\ \eqref{eq:Relative_Entropy}.

\subsection{Principle of maximum entropy}

The most prominent conceptual approach to statistical mechanics in thermal equilibrium goes through Jaynes' {\it principle of maximum entropy} \cite{Jaynes1957,Jaynes19572,Jaynes1963,Jaynes1968}. One starts with a set of {\it macroscopic} observables or state characteristics such as for example energy, particle number or magnetisation. Many {\it microscopic} quantum states might be compatible with these macroscopic characteristics. Among them, one state (i.\ e.\ a density matrix) should be preferred as having {\it maximum entropy} or {\it minimum information} by Jaynes' principle.

Why is this particular state distinguished? Take two distinct probability distributions $p$ and $q$, which both fulfill the macroscopic conditions, such that $S(p) > S(q)$. This means that the uncertainty or missing information of $p$ is greater than that of $q$ or, in other words, that $q$ contains additional information, which is not determined by the macroscopic state characteristics. Since one does not want to perform additional experiments or include information that is not available, the distribution $p$ is preferred. This is essentially the principle of minimum information or maximum entropy. 

\subsection{Principle of minimum expected relative entropy}
\label{sec:Principleofminimalrelativeentropy}

We will now attempt to formulate a principle similar to the Jaynes' maximum entropy principle, but based entirely on relative entropy. Typically, relative entropy is used to compare a model $\sigma$ with a true distribution $\rho$. 

Thus, in terms of relative entropies, a reasonable question is: What is the \textit{best model} $\sigma$ given some macroscopic state characteristics but no detailed information about the micro-state? It is important in this context that the true state $\rho$ (or the corresponding probability distribution $p$) is not known, otherwise the best model would of course be the true state itself.

The idea we will pursue in the following is to consider an average on the space of probability distributions or density matrices and to define the best model as the one that has smallest relative entropy to all possible states {\it on average}. It is then the model from which others are least distinguishable. As a prerequisite we first need to find sensible measures on the space of probability distributions and density matrices.

\subsubsection{Measure on space of probability distributions}

In the following we will first consider the {\it space of possible probability distributions} $p$ or diagonal density matrices as in eq.\ \eqref{eq:rhoDiagonal}. We will subsequently extend this to non-diagonal density matrices, as well. We want to define a sensible integral measure on this space. What is immediately clear is that the distribution should be
normalized, 
\begin{equation}
\sum_{j=1}^\mathcal{N} p_j=1.
\label{eq:normalizationp}
\end{equation}
Moreover, in practice there are typically additional constraints, such as compatibility with a set of macroscopic state characteristics. 
These constraints still leave a large degeneracy of possible micro-states, or probability distributions, of course. We construct now a {\it measure on the space of probability distributions}, which we denote by
\begin{align}
\int D p.
\label{eq:defmeasure}
\end{align}
We use here a functional integral notation because eventually we will be interested in infinite dimensional probability spaces. The set of normalized probability distributions is a manifold and one can integrate on it in terms of suitable coordinates. 
For example, the set of allowed distributions $p(\xi)$ may be parameterized by a set of parameters or coordinates $\xi = \{\xi^1, ..., \xi^m \}$, such that we can write
\begin{equation}
\int D p  = \int d \xi^1 ... \, d \xi^m \, \mu(\xi^1, ..., \xi^m).
\label{eq:DefintegralDp}
\end{equation}
We want the integral measure to be invariant under a change of coordinates $\xi \to \xi^\prime(\xi)$,
\begin{equation} 
\mu(\xi^1, ..., \xi^m) = \det\left( \frac{\partial \xi^{\prime\alpha}}{\partial\xi^\beta} \right) \mu^\prime(\xi^{\prime 1}, ..., \xi^{\prime m}).
\end{equation}

One such measure is given by Jeffreys prior as integral measure \cite{Jeffreys1946, Jaynes1968}
\begin{equation}
\mu(\xi) =\text{const} \times \sqrt{\text{det} \;  g_{\alpha\beta}(\xi)},
\label{eq:JeffreysPrior}
\end{equation}
where $g_{\alpha\beta}(\xi)$ is the Fisher metric associated with $p(\xi)$. This metric serves as Riemannian metric on the space of probability distributions and is given by \cite{Fisher1925}
\begin{equation}
\begin{split}
g_{\alpha\beta}(\xi) &= \sum_{j}  \frac{\partial  p_j (\xi)}{\partial \xi^\alpha} \frac{\partial \ln p_j (\xi)}{\partial \xi^\beta}.
\end{split}\label{eq:defFisherMetric}
\end{equation}
Based on this metric, one can define the volume form or integral measure \eqref{eq:DefintegralDp} with \eqref{eq:JeffreysPrior}. We note, however, that the measure in \eqref{eq:JeffreysPrior} is not unique. Indeed, one could multiply this by any function that is invariant under reparametrizations, such as for example $e^{-S(p\|q)}$ with some reference distribution $q$, and the measure would still be acceptable.

With the measure we just constructed one can also integrate functionals of the probability $p$, for example
\begin{equation}
\int D p \, f(p) = \int d^m \xi \, \mu(\xi) \, f(p(\xi)).
\label{eq:integralDpf}
\end{equation}
For later purpose we want to show an invariance property of expressions of this type, namely under the maps $\{ p_1, \ldots p_\mathcal{N} \} \to \{ p_{\Pi(1)}, \ldots p_{\Pi(\mathcal{N})}  \}$, where $j \rightarrow \Pi (j)$ is a permutation.\footnote{Note that this assumes a discrete probability space.} We will abbreviate this map as $p\to \Pi(p)$. The statement we want to show reads then
\begin{equation}
\int D p f(p) = \int D p f( \Pi(p)),
\label{eq:permutationInvarianceDp}
\end{equation}
for any functional $f$ of the probability distribution and any permutation $\Pi$. 

To show \eqref{eq:permutationInvarianceDp}, it is particularly convenient to parametrize the probabilities by their square roots, i.\ e.\ to write 
\begin{equation}
p_j= \begin{cases} (\xi^j)^2 & \text{for } j=1,\ldots, \mathcal{N}-1, \\ 1-(\xi^1)^2-\ldots - (\xi^{\mathcal{N}-1})^2 & \text{for } j=\mathcal{N}. \end{cases}
\label{eq:parametrization}
\end{equation}
For the Fisher metric one finds then
\begin{equation}
\frac{1}{4}g_{\alpha\beta} = \delta_{\alpha\beta} + \frac{\xi^\alpha \xi^\beta}{1-(\xi^1)^2-\ldots - (\xi^{\mathcal{N}-1})^2}.
\end{equation}
The right hand side is in fact the metric induced on the surface of a unit sphere $S_{\mathcal{N}-1}$ from the $\mathcal{N}$-dimensional Euclidean space it is embedded in \cite{Bengtsson2006, Gromov2012}.

The measure in \eqref{eq:defmeasure}, normalized to $\int Dp=1$, can be written as
\begin{equation}
\begin{split}
\int D p  = & \frac{2}{\Omega_\mathcal{N}}\int_{-1}^1 d\xi^1 \cdots d \xi^{\mathcal{N}-1} \sqrt{\det(\tfrac{1}{4} g) } \\
& \times \Theta \left(1- \sum_{\alpha=1}^{\mathcal{N}-1} (\xi^\alpha)^2\right) \\
= & \frac{1}{\Omega_\mathcal{N}}\int_{-1}^1 d \xi^1 \cdots d\xi^{\mathcal{N}} \delta\left(1- \sqrt{\sum_{\alpha=1}^\mathcal{N} (\xi^\alpha)^2}
\right).
\end{split}
\end{equation}
Here $\Omega_\mathcal{N}=2 \pi^{\mathcal{N}/2}/\Gamma(\mathcal{N}/2)$ is the surface area of the unit sphere in $\mathcal{N}$ dimensions. In this representation it is now explicit that the integral measure is invariant under permutations $p\to \Pi(p)$ as we wanted to show.

\subsubsection{Measure on space of density matrices}
Let us now extend our considerations to density matrices $\rho$. We want to integrate over all such operators that are normalized, $\text{Tr} \{ \rho \} =1$. 

The construction is very similar as for probability distributions.
We write the measure as
\begin{equation}
\int D \rho = \int d^m \xi \, \mu(\xi) = \text{const} \times \int d^m \xi \, \sqrt{\det \, g_{\alpha\beta}(\xi)},
\label{eq:measureRho}
\end{equation}
where $g_{\alpha\beta}(\xi)$ is now a Riemannian metric on the space of density matrices.
The analogue of \eqref{eq:defFisherMetric} for density matrices is the {\it quantum Fisher metric} (e.\ g.\ \cite{Braunstein1994, Facchi2010, Petz2011, Lashkari2016b, Hauke2016, Manko2017, Banerjee2018, Chruscinski2019}, see ref.\ \cite{Liu2019} for a recent overview),
\begin{equation}
g_{\alpha\beta} (\xi) = \text{Tr} \left\{ \frac{\partial \rho(\xi)}{\partial \xi^\alpha} \, \frac{\partial \ln \rho(\xi)}{\partial \xi^\beta}
\right\}.
\label{eq:DefFisherMetric}
\end{equation}
A careful consideration shows that the logarithmic derivative of a density matrix as it appears in \eqref{eq:DefFisherMetric} must be defined such that
\begin{equation}
    \frac{1}{2} \rho (d \ln \rho) + \frac{1}{2} (d \ln \rho) \rho = d\rho,
\label{eq:symLogDerivative}
\end{equation}
(it is therefore known as symmetric logarithmic derivative) and accordingly $\text{Tr}\{\rho (d \ln\rho)\}=\text{Tr}\{ d \rho \}=0$. One can also confirm from \eqref{eq:symLogDerivative} and \eqref{eq:DefFisherMetric} that $g_{\alpha\beta}(\xi)=g_{\beta\alpha}(\xi)$.

The quantum Fisher metric arises in fact as a limit of the relative entropy for density matrices that approach each other, 
\begin{equation}
    S(\rho(\xi+d\xi)||\rho(\xi)) = \frac{1}{2} g_{\alpha\beta}(\xi) d\xi^\alpha d\xi^\beta +\ldots,
    \label{eq:FisherMetricFromRelEntropy}
\end{equation}
where the ellipses stand for terms of cubic and higher order in $d\xi$. Equation \eqref{eq:FisherMetricFromRelEntropy} shows also that unitary transformations of the density matrix induce isometric transformations with respect to the quantum Fisher metric. To see this, consider a unitary map
\begin{equation}
\rho(\xi) \to \rho^\prime(\xi) = U \rho(\xi) U^\dagger = \rho(\xi^\prime).
\label{eq:UnitaryTransformRho}
\end{equation}
In the last equation we have used that $U \rho(\xi) U^\dagger$ is also a normalized density matrix, but at some other coordinate point $\xi^\prime(\xi)$. Now, from
\begin{equation}
\begin{split}
S(\rho(\xi+d\xi)||\rho(\xi)) = \, & S(U \rho(\xi+d\xi) U^\dagger || U \rho(\xi) U^\dagger) \\
 = \, & S(\rho(\xi^\prime+d\xi^\prime )||\rho(\xi^\prime)),
\end{split}
\end{equation}
and \eqref{eq:FisherMetricFromRelEntropy} it follows that $g_{\alpha\beta}(\xi) d\xi^\alpha d\xi^\beta = g_{\alpha\beta}(\xi^\prime) d\xi^{\prime\alpha} d\xi^{\prime\beta}$ so that the induced map $\xi\to \xi^\prime(\xi)$ is indeed an isometry. 

With the integral measure \eqref{eq:measureRho} one can now integrate functionals of a density matrix, similar to eq.\ \eqref{eq:integralDpf}. We want to show that this measure is invariant under unitary transformations, $\rho\to U\rho U^\dagger$, specifically,
\begin{equation}
\begin{split}
& \int D \rho \, f(\rho) = \int d^m \xi^\prime \, \mu(\xi^\prime) \, f(\rho(\xi^\prime)) \\
& = \int d^m \xi \, \mu(\xi) \, f(U \rho(\xi) U^\dagger)
= \int D \rho \,  f(U \rho U^\dagger).
\end{split}
\label{eq:invarianceU}
\end{equation}
In the second step we have used \eqref{eq:UnitaryTransformRho} together with a change of coordinates $\xi^\prime \to \xi$. 

\subsubsection{Minimizing expected relative entropy on probability distributions}

Let us now come back to the task of finding a replacement for Jaynes' principle of maximum entropy in terms of relative entropy. What we suggest is to find a state $\sigma$ that is {\it central} within the space of possible states $\rho$ in the sense it has smallest {\it expected} or {\it average} relative entropy $S(\rho||\sigma)$. It is then the state that is least distinguishable from all states, weighted by the measure or prior distribution \eqref{eq:defmeasure}, in the sense determined by relative entropy.

More concretely, when stated for diagonal density matrices or probability distribution, our {\it principle of minimum expected relative entropy} for the optimal model $q$ is to minimize the following functional
\begin{align}
B (q, \lambda) = \int D p  \left[S(p\| q) + \lambda \left( \sum_i q_i -1\right) \right].
\label{eq:BFunctional}
\end{align}
Here $\lambda$ is a Lagrange multiplier ensuring normalization of the model distribution $q$. The variation of the expression \eqref{eq:BFunctional} with respect to the model distribution $q$ can be done under the functional integral
\begin{align}
0 \overset{!}{=} \delta B = \sum_j \int D p  \left[- \frac{p_j}{q_j} + \lambda \right] \delta q_j.
\label{eq:MinRelEntropy1}
\end{align}
Fortunately the integration does not have to be executed explicitly to find an expression for $q_j$. We can simply use the invariance under permutations \eqref{eq:permutationInvarianceDp} which shows that $\int D p \, p_j$
is independent of the index $j$. This immediately implies also that the optimal distribution is independent of $j$ and must be given by the uniform distribution,
\begin{align}
q_j = \frac{1}{\mathcal{N}},
\label{eq:uniformDistribution}
\end{align}
which is equivalent to the well-known fundamental postulate of statistical physics and the microcanonical ensemble.

The argument leading to eq. \eqref{eq:uniformDistribution} is based on the permutation invariance \eqref{eq:permutationInvarianceDp} and holds for discrete probability distributions. More general, eq.\ \eqref{eq:MinRelEntropy1} is solved by $q = \langle p \rangle$ where the expectation value is with respect to the measure in eq.\ \eqref{eq:DefintegralDp}.

\subsubsection{Minimizing expected relative entropy on density matrices}

Let us now extend the variational principle \eqref{eq:MinRelEntropy1} to density matrices. To that end we compare $\rho(\xi)$ to $\sigma(\zeta)$, where $\xi$ and $\zeta$ provide convenient coordinates on the space of {\it normalized} density operators.  The functional \eqref{eq:BFunctional} gets now replaced by
\begin{equation}
B(\zeta) = \int D \rho \, S(\rho||\sigma(\zeta)) = \int d^m \xi \, \mu(\xi) \, S(\rho(\xi)||\sigma(\zeta)),
\label{eq:BFunctionalDensityOperators}
\end{equation}
and the principle of minimum expected relative entropy says that one should search for the state $\sigma(\zeta)$ corresponding to the extremum of $B(\zeta)$ in \eqref{eq:BFunctionalDensityOperators}. 

Before searching for the extremum it is convenient to decompose 
\begin{equation}
\sigma(\zeta) = U(\zeta) \, \Delta(\zeta) \, U^\dagger(\zeta), 
\end{equation}
where $\Delta(\zeta)$ is diagonal and $U(\zeta)$ is unitary. We can then write
\begin{equation}
\begin{split}
B(\zeta) = & \int D \rho \, S(\rho|| U(\zeta)  \Delta(\zeta)  U^\dagger(\zeta) ) \\
= & \int D \rho \, S( U^\dagger(\zeta) \rho U(\zeta)|| \Delta(\zeta) ) =   \int D \rho \, S(\rho|| \Delta(\zeta) ). 
\end{split}
\label{eq:invarianceBU}
\end{equation}
In the second step we have used invariance of relative entropy under unitary transformations and in the last step the invariance property of the functional integral, eq.\ \eqref{eq:invarianceU}.  Eq.\ \eqref{eq:invarianceBU} shows that the functional $B(\zeta)$ is in fact degenerate for all density matrices $\sigma(\zeta)$ that are related through unitary transformations and depends therefore only on the eigenvalues of $\sigma(\zeta)$. 

The right hand side of eq.\ \eqref{eq:invarianceBU} is convenient for variation,
\begin{equation}
0 \overset{!}{=} \delta B =  \sum_j \int D \rho \left[ - \frac{ \rho_{jj}}{\Delta_{jj}} +\lambda \right] \delta \Delta_{jj}.
\end{equation}
For simplicity we took now a parametrization in terms of the diagonal values $\Delta_{jj}$ themselves and introduced a Lagrange multiplier to ensure normalization, similar as in eq.\ \eqref{eq:MinRelEntropy1}. 
Interestingly, the square bracket on the right hand side depends only on the diagonal elements of the density operator $\rho$. One can now use that the diagonal elements $\rho_{jj}$ can be permuted, $\rho_{jj} \to \rho_{\Pi(j) \Pi(j)}$, by special unitary transformations\footnote{An example for two-dimensional matrices is $U=i\sigma_2$.} $\rho\to U \rho U^\dagger$, and therefore it follows from eq.\ \eqref{eq:invarianceU} that $\int D\rho \, \rho_{jj}$ is again independent of the index $j$, similar as we had it for probability distributions. The extremum is then the uniform density matrix
\begin{equation}
\sigma_\text{m} = \frac{1}{\mathcal{N}} \mathbbm{1}.
\label{eq:uniformDensityMatrix}
\end{equation}
Incidentally, in light of \eqref{eq:invarianceBU} this is anyway the only possibility to have a unique extremum because it is the only density matrix that remains unchanged by all unitary transformations $\sigma \to U \sigma U^\dagger$.

To summarize, the uniform distribution \eqref{eq:uniformDistribution} and density matrix \eqref{eq:uniformDensityMatrix} can also be obtained through a variational principle that is entirely based on relative entropy instead of entropy. We now close this subsection with a simple example.

\subsubsection{Example: 2-state system}

\begin{figure*}[t!]
    \centering
    \includegraphics[height=0.2\textwidth]{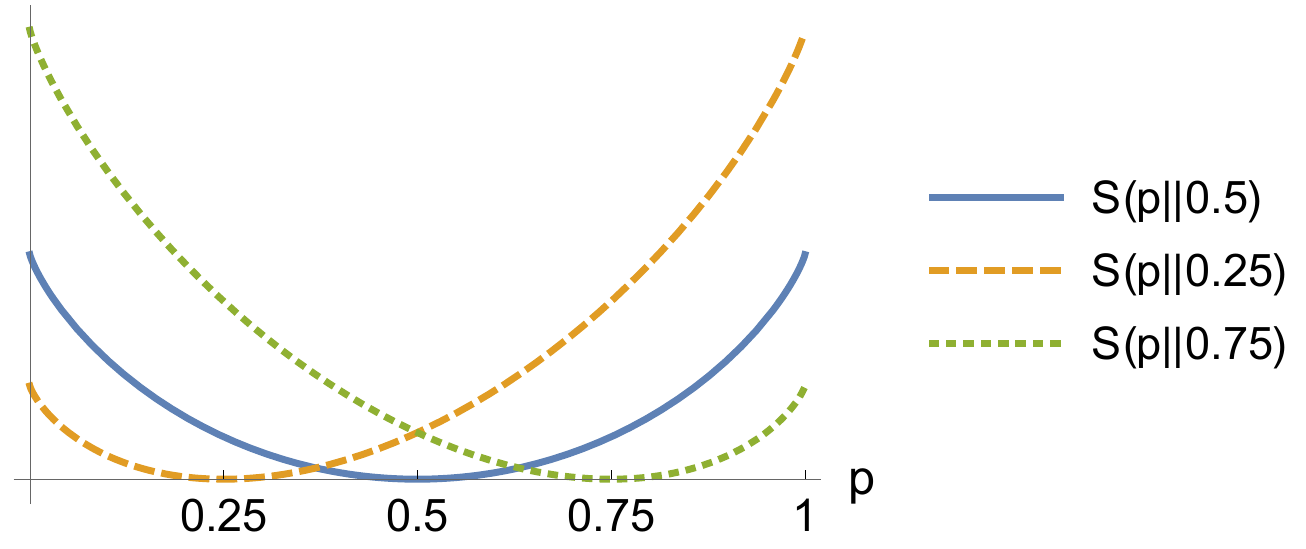}
    \hfill
    \includegraphics[height=0.2\textwidth]{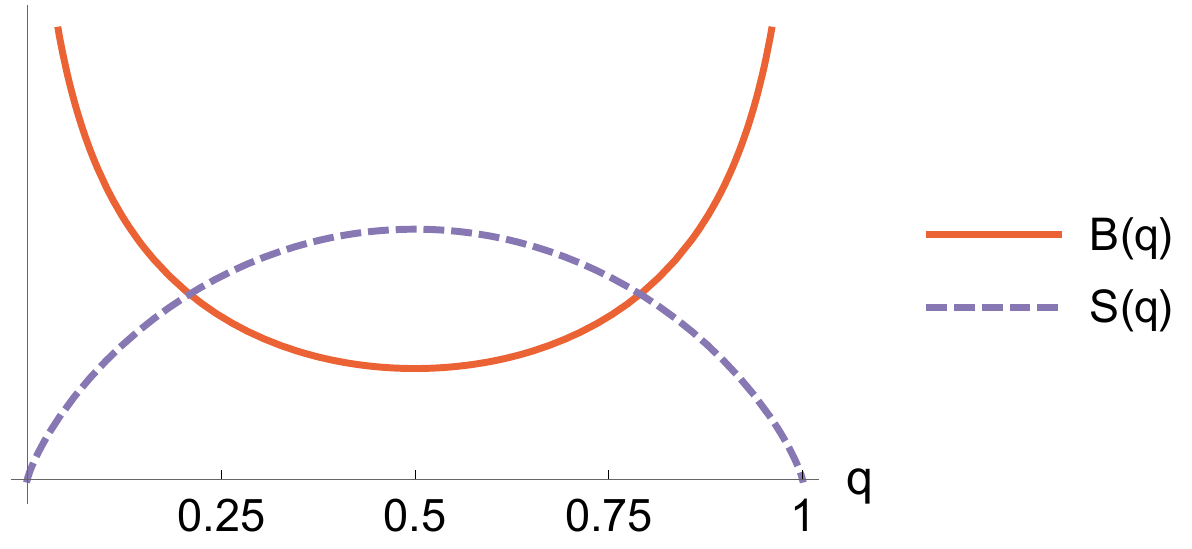}
    \caption{Left panel: Curves of relative entropies $S(p\|q)$ for a two-state system with probabilities $p_1=p$, $p_2=1-p$ with respect to  different model distributions $q_1=q$, $q_2=1-q$ as parametrized by the value of $q$. Right panel: Entropy $S(q)$ and area function $B(q)$ of the model $q$. The maximum of entropy and the minimum of $B(q)$ are both at the uniform distribution, $q=1/2$.} 
    \label{fig:Entropies2State}
\end{figure*}

Here we want to present the explicit form of the functional $B(q,\lambda)$ in eq.\ \eqref{eq:BFunctional} for a simple example, the 2-state system. Then the parametrization \eqref{eq:parametrization} reads $p_1 = \xi^2$ and $p_2 = 1-\xi^2$, which leads to the Fisher information
\begin{equation}
    \frac{1}{4} g(\xi) = 1 + \frac{\xi^2}{1 - \xi^2}.
\end{equation}
At the extremum with respect to $\lambda$, the distribution $q_1=q$, $q_2=1-q$ is normalized and we obtain for the remaining function $B(q)$ the simple expression
\begin{equation}
    B(q) = 1 - 2 \ln 2 - \frac{1}{2} \ln \left(q (1-q) \right).
\end{equation}
But what is the interpretation of this function? Let's have a look at relative entropies with respect to different models first (left panel of \autoref{fig:Entropies2State}). If we consider the best model the resulting relative entropy curve takes its minimum at $p=\frac{1}{2}$. For other models one gets asymmetric curves. In this simple case $B(q)$ is actually the area under the curve given by the corresponding relative entropy with respect to a fixed model $q$. One can directly see that the area under a relative entropy curve is minimized for $q=\frac{1}{2}$ and goes to infinity for $q \rightarrow 0$ and $q \rightarrow 1$, which is also visible in the right panel of \autoref{fig:Entropies2State}. Thus the area function is indeed minimized where the entropy of the corresponding model is maximized.

\subsection{Updating knowledge with expectation values}

The principle of minimum expected relative entropy introduced above is enough to describe besides the microcanonical ensemble also the canonical and the grand-canonical ensemble. We will discuss this in more detail in section \ref{chap:Thermodynamics}. 

Here we want to mention for completeness also a related principle used in statistical inference, typically to update an ensemble with additional knowledge in the form of expectations values. This is the {\it principle of minimum discrimination information} or {\it principle of minimum cross entropy} \cite{Kullback1959, Cover2006}. It has many applications in the context of Monte Carlo simulations, optimisation problems and machine learning \cite{Rubinstein2004, Rubinstein2005}.

The idea (translated to physics) is as follows. Suppose we have found some model probability distribution $q$ for some physical situation. Then somebody gives us additional information in the form of $k$ expectations values, which is not already taken into account by our current model $q$. Now we want to improve our model by implementing this additional information, which should lead to a new model distribution~$p$. 

It is now sensible to do this with minimal gain of knowledge, or, in terms of relative entropy, through minimizing $S(p\|q)$ with respect to $p$ under $k+1$ constraints (the updated distribution has to be normalized as usual, the prior distribution is normalized by construction).

To find a general expression for the probability distribution $p$ we use again the technique of Lagrange multipliers. Let's say we have observables $A_j$ where $j \in \{1,...,k \}$, where each observable $A_{j}$ has values $a_{j,i}$ in microstates with index $i \in \{1,...,\mathcal{N} \}$. Then we assign $k$ multipliers $\lambda_j$ to the $k$ expectation values and add one multiplier $\gamma$ for normalization. The Lagrange function is
\begin{equation}
\begin{split}
\mathcal{L} (p,q,\lambda_j,\gamma) = & \, S(p \| q) + \sum_{j=1}^{k} \lambda_j \left( \sum_{i=1}^{\mathcal{N}} p_i a_{j,i} - A_j \right) \\ 
&+ \gamma \left(\sum_{m=1}^{n} p_m - 1 \right).
\end{split}
\end{equation} 
We want to compute the minimum 
\begin{align}
\delta \mathcal{L} (p,q,\lambda_j,\gamma) \overset{!}{=} 0,
\end{align}
where the variation is now with respect to the improved distribution $p$. 

Solving the above extremization problem leads to
\begin{align}
p_i = \frac{q_i}{\bar{Z}} \exp \left(- \sum_{j=1}^{k} \lambda_j a_{j,i} \right),
\label{eq:Boltzmann_Probabilities}
\end{align}
where $\bar{Z}$ ensures normalization. So if we start with a model $q$ and add information from expectation values, we get an improved model $p$ which is effectively the prior model $q$ times weight factors enforcing expectation values $A_j$.

One notes the resemblance between \eqref{eq:Boltzmann_Probabilities} and Boltzmann-type probability weights $e^{-\beta E}$. However, the superficial impression of such a connection might be a bit misleading. Specifically, if one tries to use the principle of minimum discrimination power following the above discussion to introduce the canonical ensemble with probabilities $p_i = \frac{1}{Z}e^{-\beta E_i}$, one needs to start with prior probabilities $q_i$ that are flat with respect to energy. This does not describe any physically sensible state for most systems -- formally it corresponds to a limit of the canonical ensemble with infinite temperature. 

There might be other situations, however, where the principle of minimal discrimination information can be applied to physics problems. As an example, let us consider a single spin as part of a larger spin system which is in equilibrium with a heat bath of temperature $T$. We take the magnetic moment of the spin to be $\vec \mu_i = \gamma \vec s_i$ in the state $i$.
We may consider a ferromagnetic system at vanishing {\it external} magnetic field $H=0$. Starting point is a distribution of states according to Boltzmann weights 
\begin{equation}
q_i = \frac{1}{\tilde Z}e^{-\beta E_i}. 
\end{equation}
This serves as a prior model in the following. Without further specifications, all spin directions $\vec s_i$ are equally likely. However, a ferromagnetic state has domains with non-vanishing magnetization. Concentrating now on one such region with given magnetization $\vec M$, one may introduce a Lagrange multiplier $-\beta \vec B$ for the magnetic moment $\vec \mu$ and the probabilities become
\begin{equation}
p_i = \frac{q_i}{\bar Z} e^{\beta \gamma \vec B \vec s_i} = \frac{1}{Z} e^{-\beta(E_i-\gamma \vec B \vec s_i)},
\end{equation}
where $Z=\tilde Z \bar Z=\sum_i e^{-\beta(E_i-\gamma \vec B \vec s_i)}$. This updates the model distribution for the single spin. Proceeding similarly with other spins would lead to a mean-field type description. There the in-medium magnetic field $\vec B$ would have to be determined in a self-consistent way.

\section{Thermodynamics}
\label{chap:Thermodynamics}

In this section we will use relative entropy and the principle of minimum expected relative entropy to discuss different thermodynamical ensembles as well as the definition of temperature and chemical potential. Eventually we will also formulate the third law of thermodynamics in terms of relative entropy.

\subsection{Microcanonical ensemble}

We consider a situation with fixed energy $E$ and particle number $N$. (Here and below we will always keep the volume $V$ fixed.) We have already seen in section \ref{sec:Principleofminimalrelativeentropy} that the principle of minimum expected relative entropy (on the space of states in agreement with the constraints) leads to the uniform distribution or microcanonical density matrix 
\begin{equation}
\sigma_{\text{m}} = \frac{1}{Z_\text{m}} \delta(H-E(\sigma_{\text{m}})) \delta (N-N(\sigma_{\text{m}})),
\label{eq:microcanonicalDensityMatrix}
\end{equation}
where
\begin{equation}
Z_\text{m} = \text{Tr} \{ \delta(H-E(\sigma_{\text{m}})) \delta (N-N(\sigma_{\text{m}})) \}.
\end{equation}

We are interested in computing the relative entropy of an arbitrary state $\rho$ to the microcanonical state \eqref{eq:microcanonicalDensityMatrix},
\begin{align}
S(\rho \| \sigma_{\text{m}}) &= \text{Tr} \{ \rho \ln \rho - \rho \ln \sigma_{\text{m}} \} \notag \\
&= - S(\rho) - \text{Tr} \{ \rho \ln \sigma_{\text{m}} \}.
\end{align}
The first term is simply the negative von Neumann entropy of $\rho$ while the second is known as the cross entropy. Using that $\sigma_{\text{m}}$ is constant in the subspace of states with energy $E(\sigma_{\text{m}})$ and particle number $N(\sigma_{\text{m}})$ where it has support, and that both states are normalized, leads for $\text{supp}(\rho) \subseteq \text{supp}(\sigma_\text{m})$ to
\begin{align}
-\text{Tr} \{\rho \ln \sigma_{\text{m}} \} = -\text{Tr} \{\sigma_{\text{m}} \ln \sigma_{\text{m}} \} = S (\sigma_{\text{m}}).
\end{align}
The support condition for $\rho$ translates to a condition on the fixed values of energy and particle number. Both states have to have the same energy and particle number, otherwise the support condition is violated and their relative entropy becomes infinite. Note that it is not enough if $\rho$ has an energy expectation value $\langle H \rangle_\rho = \text{Tr}\{ \rho H \}$ that agrees with $E(\sigma_\text{m})$; also the dispersion needs to vanish so that $\langle H ^2 \rangle_\rho = \text{Tr}\{ \rho H^2 \}=E(\sigma_\text{m})^2$. We denote this strict condition by $E(\rho)\equiv E(\sigma_\text{m})$ and similarly for particle number $N(\rho) \equiv N(\sigma_\text{m})$. In summary one finds
\begin{align}
S (\rho \| \sigma_{\text{m}}) = \begin{cases}
  - S (\rho) + S (\sigma_{\text{m}}) &\text{for } E(\rho) \equiv E(\sigma_{\text{m}}) \\ & \text{and } N(\rho) \equiv N(\sigma_{\text{m}}), \\
 +\infty &\text{else}.
\end{cases}
\label{eq:MicroEnsembleRelEntropy}
\end{align}
In other words, if we model a state $\rho$, which has a definite energy and particle number, with a microcanonical state $\sigma_{\text{m}}$ with the same energy and particle number, then their relative entropy is just the difference of entropies. If the energies or particle numbers do not agree, the relative entropy becomes infinite. Intuitively, in that case it is rather easy to distinguish the states, because a measurement of $E$ or $N$ is sufficient.

As shown in section \ref{sec:Principleofminimalrelativeentropy}, the microcanonical state $\sigma_\text{m}$ corresponds to the state to which other states $\rho$ have on average the smallest relative entropy. Equation \eqref{eq:MicroEnsembleRelEntropy} tells that this relative entropy can then be written as a difference of entropies.
Moreover non-negativity of relative entropy highlights that $\sigma_{\text{m}}$ is indeed the state with maximum entropy for given $E$ and $N$, because $S(\rho \| \sigma_{\text{m}}) \ge 0$ is equivalent to
\begin{equation}
 S(\rho) \le S(\sigma_{\text{m}})
\label{eq:Relative_Entropy_Micro},
\end{equation}
for all states $\rho$ with $E(\rho)\equiv E(\sigma_{\text{m}})$ and $N(\rho) \equiv N(\sigma_{\text{m}})$.

It is instructive to consider also the differential of \eqref{eq:MicroEnsembleRelEntropy} for $dE(\rho) \equiv dE(\sigma_\text{m})$ and $dN(\rho) \equiv dN(\sigma_\text{m})$,
\begin{equation}
\begin{split}
d S (\rho \| \sigma_{\text{m}}) = & - dS(\rho) + dS(\sigma_\text{m}), \\
= & - dS(\rho) + \beta \, dE(\rho) - \beta \mu \, dN(\rho).
\end{split}
\label{eq:differentialSrelamicro}
\end{equation}
In the second step we used the standard thermodynamic relation
\begin{equation}
dS(\sigma_\text{m}) = \beta \, dE(\sigma_\text{m}) - \beta \mu \, dN(\sigma_\text{m}).
\end{equation}
Interestingly, \eqref{eq:differentialSrelamicro} provides a possibility to define the inverse temperature $\beta=1/T$ and chemical potential $\mu$ of a microcanonical state $\sigma_\text{m}$ from partial derivatives of a relative entropy, at fixed entropy $S(\rho)$. It is important to fulfill $dE(\rho) \equiv dE(\sigma_\text{m})$ and $dN(\rho) \equiv dN(\sigma_\text{m})$, though.

\subsection{Thermal fluctuations}

Let us discuss here also briefly the meaning of relative entropy for the description of thermal fluctuations. Even in a fully equilibrated state, quantities can differ from their thermal equilibrium expectation values, for example the energy or particle number in a subvolume. Typically, the relative importance of fluctuations is larger when the subvolume or subsystem they concern is smaller. The traditional theory of such thermodynamic fluctuations goes back to Einstein's work on critical opalescence \cite{Einstein1910} (see also ref.\ \cite{Landau1980} and for a recent discussion ref.\ \cite{Mishin2015}) and one associates to a fluctuation of a macroscopic variable $\xi$ a probability proportional to $e^{S(\xi)}$ where $S(\xi)$ is the entropy. The quantity $\xi$ is here such that it does not have a sharp value but fluctuates in the thermal equilibrium state $\sigma$. The entropy $S(\xi)$ can be understood as the entropy of an ensemble of microstates, or a density matrix $\rho(\xi)$, for which the macroscopic variable $\xi$ has a fixed value but that is otherwise in agreement with conservation laws and possibly other relevant constraints. The entropy $S(\xi)=S(\rho(\xi))$ is strictly smaller than the equilibrium entropy $S(\sigma)$ because the latter is maximal (within the constraints) and because $\rho(\xi)$ is necessarily different from the equilibrium density matrix $\sigma$. One can take the probability distribution for a fluctuation to be proportional to $e^{S(\rho(\xi)) - S(\sigma)}$. We note here already the close connection to relative entropy in eq.\ \eqref{eq:MicroEnsembleRelEntropy}. 

As we have discussed in section \ref{chap:Introduction}, in (classical) statistical interference, $e^{-N S(p\|q)}$ governs the asymptotic probability to find after $N$ drawings relative frequencies $p_j=N(x_j) / N$ for a random variable $x_j$ that is actually distributed according $q_j$, see eq.\ \eqref{eq:relativeEntropyForFluctuation} and the discussion there. This can be understood as a fluctuation of the frequency $p_j$ around the expectation value $\langle p_j \rangle = q_j$. 
This setup is actually closely related to the one of thermal fluctuations. If the number of drawings $N$, or the size of the relevant subsystem in the case of thermal fluctuations, grows, the probability for sizeable fluctuations (relative to the expectation value) becomes quickly very small.

Note, however, that eq.\ \eqref{eq:relativeEntropyForFluctuation} describes the asymptotic limit of many drawings $N\to \infty$, while thermal fluctuations concern in some sense finite size corrections to the thermodynamic limit of a subsystem. Nevertheless, the relation motivates why the probability for finding a state $\rho(\xi)$ in agreement with a macroscopic value $\xi$ should be proportional to $e^{-S(\rho(\xi)\|\sigma_\text{m})}$, when $\sigma_\text{m}$ is the actual thermal state. Using \eqref{eq:MicroEnsembleRelEntropy}, we see that this indeed agrees with the traditional theory of thermal fluctuations based on entropy. Note that we can formally also allow $\rho$ to violate the conditions $E(\rho)\equiv E(\sigma_\text{m})$ and $N(\rho)\equiv N(\sigma_\text{m})$; the probability for such fluctuations then simply vanishes.

To achieve reparametrization invariance, one must use the invariant volume element in parameter space and one can state that the probability for a thermal fluctuation in the parameter volume $d^m \xi$ should be
\begin{equation}
dW = \frac{1}{Z} e^{-S(\rho(\xi) \|\sigma_\text{m})} \sqrt{\text{det} \, g_{\alpha\beta}(\xi)} \, d^m \xi,
\label{eq:thermalFluctuationDistribution}
\end{equation}
where $g_{\alpha\beta}(\xi)$ is the Fisher metric for $\rho(\xi)$ as defined in eq.\ \eqref{eq:DefFisherMetric} and $Z$ is defined such that $\int dW=1$. 

To make things more concrete, let us assume that the exponential term in \eqref{eq:thermalFluctuationDistribution} has a maximum at $\xi_0$. Let us assume moreover, that one can approximate\footnote{This can not be exact because $\xi$ is fluctuating for $\sigma$ but fixed for $\rho(\xi_0)$.} $\rho(\xi_0)\approx\sigma_\text{m}$ and consider a quadratic approximation (in $\xi-\xi_0$) to the relative entropy. We have then
\begin{equation}
dW = \frac{1}{Z} e^{-\frac{1}{2} g_{\mu\nu}(\xi-\xi_0)^\mu (\xi-\xi_0)^\nu} \sqrt{\text{det}(g)}\, d^m \xi,
\end{equation}
where the Fisher information metric appears now also in the exponent. Here we can easily determine $Z=(2\pi)^{m/2}$. Thermal fluctuations are in a Gaussian approximation directly determined by the Fisher information metric of the equilibrium distribution. More generally, thermal fluctuations are governed by relative entropy, though.

Before we close this subsection, let us remark that thermal fluctuations can also be described in the canonical and grand canonical ensemble by relative entropy. The density matrix of the microcanonical ensemble $\sigma_\text{m}$ in eq.\ \eqref{eq:thermalFluctuationDistribution} is then simply replaced by the density matrix of the canonical ensemble $\sigma_\text{c}$ or grand-canonical ensemble $\sigma_\text{gc}$, respectively. One can confirm that the relative entropies in eqs.\ \eqref{eq:Relative_Entropy_Canonical} and \eqref{eq:Relative_Entropy_GrandCanonical} lead to the same expressions as the traditional theory formulated with entropy (see for example ref.\ \cite{Landau1980}).

\subsection{Canonical ensemble}

The transition from the micro-canonical to the canonical ensemble can be done by following essentially the usual construction. Suppose we still have an overall isolated system with fixed total energy $E$ and fixed total particle number $N$, but we consider a decomposition into a small subsystem $A$ and a heat bath $B$. Furthermore the two subsystems are allowed to exchange energy with each other. The question is now: How can we deduce the \textit{best model} $\sigma_{\text{c}}$ for the small system without using secondary fundamental principles? 

The subsystem $A$ may have an energy $E_A$ while $B$ has then energy $E_B = E-E_A$. However, this decomposition is not fixed, but fluctuating. Equilibration between $A$ and $B$ needs an interaction between them. However, for the equilibrium state itself, the role of this interaction term is typically neglected and one assumes that the Hamiltonian can be decomposed into two independent terms, $H=H_A + H_B$. For the microcanonical state we can then write
\begin{equation}
\sigma_\text{m} \sim \int_0^E d E_A \frac{dW}{dE_A} \delta(H_A - E_A) \delta (H_B - E + E_A).
\end{equation}
We use here the distribution $dW/dE_A$ with $\int dW = 1$ where all values for $E_A$ contribute that are allowed by overall energy conservation. Note in particular that $\sigma_\text{m}$ is {\it not} a product state.

It is nevertheless instructive to consider a class of product states in the form
\begin{equation}
\rho(E_A) = \rho_A(E_A) \otimes \rho_B(E-E_A),
\end{equation}
and specifically the relative entropy (using eq.\ \eqref{eq:MicroEnsembleRelEntropy})
\begin{equation}
S(\rho(E_A) \| \sigma_\text{m}) = - S(\rho_A(E_A)) - S(\rho_B(E-E_A)) + S(\sigma_\text{m}).
\label{eq:relativeEntropyTwoSystemsEnergy}
\end{equation}
Because relative entropy is positive, there needs to be a minimum of \eqref{eq:relativeEntropyTwoSystemsEnergy}. The two negative terms on the right hand side are maximized when $\rho_A(E_A)$ and $\rho_B(E-E_A)$ are microcanonical density matrices in the respective subspaces. Searching then for a minimum by variation with respect to $E_A$ leads to the usual condition
\begin{equation}
\frac{\partial}{\partial E} S(\rho_A(E)){\big |}_{E_A} =  \frac{\partial}{\partial E} S(\rho_B(E) ){\big |}_{E-E_A}.
\end{equation}
The inverse temperature in both subsystems needs to agree. The latter is given by the usual relation
\begin{equation}
\beta = \frac{\partial}{\partial E} S(\rho(E)).
\end{equation}

One may now determine a density matrix for the subsystem $A$. This can be based on eq.\ \eqref{eq:thermalFluctuationDistribution} giving the probability for thermal fluctuations. The energy of the subsystem $E_A$ plays here the role of a coordinate $\xi$. Using \eqref{eq:relativeEntropyTwoSystemsEnergy} and expanding the relative entropy to linear order in $E_A$
leads to
\begin{equation}
S(\rho(E_A) \| \sigma_\text{m}) = %- S(\rho_A(E_A)) + 
\beta E_A + \text{const}.
\label{eq:relEntropyExpansionCan}
\end{equation}
This approximation can now be inserted in eq.\ \eqref{eq:thermalFluctuationDistribution} and $E_A$ can be replaced by the Hamiltonian of subsystem $A$. One finds that the density matrix for $A$ must be of the canonical form
\begin{equation}
\sigma_c = \frac{1}{Z} e^{-\beta H},
\label{eq:canonicalDensityMatrix}
\end{equation}
with normalization factor
\begin{equation}
Z = e^{-\beta F} = \text{Tr} \{ e^{-\beta H} \}.
\end{equation}
Both the Hamiltonian $H$ and the trace $\text{Tr}$ are now restricted to system $A$. Because of the possibility of energy exchange, states of different energies can now appear, but they are weighted by Boltzmann factors. 

In a next step, let us now determine the relative entropy of an arbitrary state $\rho$ with respect to the canonical density matrix \eqref{eq:canonicalDensityMatrix}. One finds
\begin{align}
S(\rho \| \sigma_{\text{c}}) &= - S (\rho) - \text{Tr} \{ - \rho \ln Z - \rho \beta H \} \notag \\
&= - S(\rho) + \ln Z + \beta E(\rho),
\end{align}
where $E(\rho) = \braket{H}_\rho=\text{Tr}\{ \rho H \}$ was used. To evaluate the logarithm of the partition function we can use the relations
\begin{align}
    E(\sigma_{\text{c}}) &= - \frac{\partial}{\partial \beta} \ln Z, \\
    S(\sigma_{\text{c}}) &=- \beta F (\sigma_c) + \beta E(\sigma_\text{c}) = \ln Z - \beta \frac{\partial}{\partial \beta} \ln Z.
\end{align}
Combining them leads to
\begin{align}
    \ln Z = -\beta E(\sigma_{\text{c}}) + S(\sigma_{\text{c}}),
\end{align}
such that we end up with
\begin{equation}
S(\rho \| \sigma_{\text{c}}) = \begin{cases}  - S(\rho) +  S(\sigma_{\text{c}}) + \beta \hspace*{-0.3cm}& \left( E(\rho) - E(\sigma_{\text{c}}) \right) \\ & \text{for } N(\rho)\equiv N(\sigma_\text{c}), \\ \infty & \text{else}.\end{cases}
\label{eq:Relative_Entropy_Canonical}
\end{equation}
We see that, compared to the microcanonical model in eq.\ \eqref{eq:MicroEnsembleRelEntropy}, the relative entropy acquires a second term due to a possible difference in energies. Furthermore the support condition is released somewhat. Only if the two particle numbers do not agree, the relative entropy is still infinite. Moreover we get back the microcanonical result \eqref{eq:Relative_Entropy_Micro} if the energy expectation values match.

The relative entropy between the state $\rho$ and the canonical state $\sigma_{\text{c}}$ also has a very intuitive physical meaning (if the support condition is met). Up to a factor $\beta$ it describes the so-called \textit{available energy}, which is the maximum work, that can be extracted from the system being in the state $\rho$, if it is in contact with a heat bath with inverse temperature $\beta$ \cite{Hatsopoulos1976I,Hatsopoulos1976IIa,Hatsopoulos1976IIb,Hatsopoulos1976III}. In the special case where $\rho=\rho_\text{c}$ is an equilibrium state with an in general different inverse temperature $\beta'$, the available energy reduces to a difference of free energies.

It is instructive to consider the differential of the relative entropy \eqref{eq:Relative_Entropy_Canonical} (restricted to $dN(\sigma_c)\equiv dN(\rho)$),
\begin{equation}
\begin{split}
   d S(\rho \| \sigma_{\text{c}}) = & - d S(\rho) + d S(\sigma_{\text{c}}) + (E(\rho)-E(\sigma_{\text{c}})) \, d\beta \\ 
    & + \beta \left( d E(\rho) -  d E(\sigma_{\text{c}}) \right) \\
    = & - d S(\rho) + \beta \, d E(\rho) - \beta \mu \, dN(\rho) \\
    & + (E(\rho)-E(\sigma_{\text{c}})) \, d \beta,
\end{split}    
\label{eq:RelEntropy_TotalDiff_Explicit}
\end{equation}
where we have used that for the thermal state
\begin{align}
    d S(\sigma_{\text{c}}) = \beta  d E(\sigma_{\text{c}}) - \beta \mu \, dN(\sigma_c).
\end{align}
If we consider $S(\rho)$, $E(\rho)$, $N(\rho)$ and the inverse temperature $\beta$ of the state $\sigma_\text{c}$ as independent, we can read off from \eqref{eq:RelEntropy_TotalDiff_Explicit} that 
\begin{equation}
\begin{split}
     & \frac{\partial S(\rho \| \sigma_{\text{c}})}{\partial S(\rho)} {\Big |}_{E(\rho), N(\rho), \beta} = -1, \\
     & \frac{\partial S(\rho \| \sigma_{\text{c}})}{\partial E(\rho)}  {\Big |}_{S(\rho), N(\rho), \beta} = \beta, \; \frac{\partial S(\rho \| \sigma_{\text{c}})}{\partial N(\rho)}  {\Big |}_{S(\rho), E(\rho), \beta}= - \beta \mu,\\
     & \frac{\partial S(\rho \| \sigma_{\text{c}})}{\partial \beta}  {\Big |}_{S(\rho), E(\rho), N(\rho)}= E(\rho) - E(\sigma).
\end{split}
\end{equation}
Let's have a closer look on these relations. The first relation tells us that if the information about the true state $\rho$ increases (which corresponds to a decreasing entropy), the relative entropy with respect to the thermal state increases in the same way. This is intuitively clear since the thermal state is the state representing maximal missing information and the actual state veers away from the thermal state if its information increases. 

The relations in the second line provide an interesting possibility to  define temperature and chemical potential of a canonical density matrix $\sigma_\text{c}$ through partial derivatives of a relative entropy at fixed $S(\rho)$. 

The relation in the third line has also an interesting meaning. We observe that precisely if we choose the two energy expectation values to be equal, the partial derivative with respect to $\beta$ vanishes,
\begin{align}
    \frac{\partial S(\rho \| \sigma_{\text{c}})}{\partial \beta} = 0 \quad\quad \Leftrightarrow \quad\quad E(\rho) = E(\sigma).
\end{align}
In other words, if we choose the model energy to be the correct energy, the relative entropy is minimized with respect to the inverse temperature. Since the temperature is the only degree of freedom of a canonical state, this choice corresponds to the optimal canonical model; all other choices lead to a larger relative entropy.

While we have obtained here the canonical ensemble from the microcanonical ensemble using the standard procedure, one may also ask whether a principle of minimum expected relative entropy can be formulated directly in the canonical case. Indeed this could be done similarly to the construction in section \ref{sec:Principleofminimalrelativeentropy}, however the integral $Dp$ over probability distributions, or $D\rho$ over density matrices, would now have to go over a set of states that does not have fixed energy. The detailed properties of the appropriate functional integral measure will have to be fixed in more detailed investigations beyond the scope of the present work. However, there is one interesting statement one can make on rather general terms, generalizing the discussion in section \ref{sec:Principleofminimalrelativeentropy}. 

Consider a functional as in eq.\ \eqref{eq:BFunctionalDensityOperators} but with a more general definition of the functional integral measure $D\rho$. We are interested in finding the minimal expected relative entropy and vary therefore with respect to the density matrix $\sigma$,
\begin{equation}
\delta B = \int D\rho \; \text{Tr}\{ \rho \, d \ln \sigma \} = \text{Tr} \left\{ \langle \rho \rangle \, d \ln \sigma \right\}.
\end{equation}
In the second equation we use $\langle \cdot \rangle = \int D\rho (\cdot)$ for the expectation value with respect to the functional integral over $\rho$. Our claim is now that $B$ is stationary for $\sigma = \langle \rho \rangle$. To see this we first use the cyclic property of the trace to write
\begin{equation}
\delta B = \frac{1}{2}\text{Tr} \left\{ \langle \rho \rangle \, (d \ln \sigma )+ (d \ln \sigma) \, \langle \rho \rangle \right\}.
\end{equation}
The logarithmic derivative should be understood here as a symmetric logarithmic derivative as defined in \eqref{eq:symLogDerivative}. One can then see that for $\sigma = \langle \rho \rangle$ one has 
\begin{equation}
\delta B = \text{Tr} \{ d \sigma \} = 0.
\end{equation}
In the last step we used that $\text{Tr}\{\sigma\}=1$ needs to be normalized.

These considerations show that one can obtain the canonical density matrix directly from a principle of minimum expected relative entropy when the integral measure $D\rho$ is such that the expectation values is the canonical density matrix, $\langle \rho \rangle = \sigma_c$.

\subsection{Grand canonical ensemble}

The same analysis can be done for a small system in contact with a heat and particle bath. There we expect the best model for the small system to be the grand canonical ensemble. We abbreviate the technical steps and directly use the well-known expression
\begin{align}
\sigma_{\text{gc}} = \frac{1}{Z} e^{-\beta (H -\mu N)},
\end{align}
where $\mu$ is the chemical potential, allowing for particle exchange between the two systems, and $Z$ is now the grand canonical partition sum. For the relative entropy of an arbitrary state $\rho$ and the grand canonical model one finds
\begin{align}
S(\rho \| \sigma_{\text{gc}}) = & - S(\rho) + S(\sigma_{\text{gc}}) + \beta \left( E(\rho) - E(\sigma_{\text{gc}}) \right) \notag \\ & -  \beta \mu \, (N(\rho) - N(\sigma_{\text{gc}})).
\label{eq:Relative_Entropy_GrandCanonical}
\end{align}
We see that the full relation is linear in differences of extensive quantities, as may be expected from the canonical case. Moreover the support condition is released even further. The only case, in which the relative entropy would still be infinite is if the two states would live in different volumes (we have excluded this possibility so far).

The differential of the relative entropy \eqref{eq:Relative_Entropy_GrandCanonical} becomes now
\begin{equation}
\begin{split}
   d S(\rho \| \sigma_{\text{gc}}) = & - d S(\rho) + \beta \,d E(\rho) - \beta\mu \, d N(\rho) \\
    &+(E(\rho)-E(\sigma_{\text{gc}})) \, d \beta  \\
    &- (N(\rho)-N(\sigma_{\text{gc}})) \, d(\beta\mu),
\end{split}
\end{equation}
which implies for example the relations 
\begin{align}
    \frac{\partial S(\rho \| \sigma_{\text{gc}})}{\partial N(\rho)} {\Big |}_{S(\rho), E(\rho), \beta, \mu} &= - \beta\mu, \\
    \frac{\partial S(\rho \| \sigma_{\text{gc}})}{\partial (\beta\mu)} {\Big |}_{S(\rho), E(\rho), N(\rho), \beta} &= - N(\rho) + N(\sigma_{\text{gc}}).
\end{align}
These two relations have an analogous interpretation as before. The optimal choices of $\beta$ and $\mu$, in the sense of a stationary relative entropy correspond to choosing a model with coinciding expectation values for energy, $E(\rho) = E(\sigma_\text{gc})$, and particle number, $N(\rho) = N(\sigma_{\text{gc}})$.

\subsection{Third law of thermodynamics}
Entropy does not only appear as a thermodynamic potential, but is also central for the three laws of thermodynamics. Of special interest here is the third law. This is because the first law expresses simply energy conservation, which does not have to be linked to entropy necessarily and the second law was formulated in several ways using relative entropy already (for an overview see \cite{Sagawa2012} and also \cite{Haas2020}).

There exist several possible formulations of the third law. A popular approach is Planck's formulation: Entropy approaches a constant value for the temperature going to zero, $T \rightarrow 0$, independently of all other thermodynamic parameters. A quantum mechanical interpretation allows to identify this constant with the entropy $S_0$ of the ground state $\rho_0$, because the state of the system $\rho$ approaches the ground state $\rho_0$ as temperature decreases. If we model the actual state $\rho$ by either a canonical or grand-canonical thermodynamic model $\sigma$ we may formulate the third law as follows.

\paragraph*{Third law of thermodynamics:} The relative entropy $S(\rho_0 \| \sigma)$ between the ground state $\rho_0$ and a thermodynamic model state $\sigma$ approaches zero for $T \rightarrow 0$,
\begin{align}
    \lim_{T \to 0} S(\rho_0 \| \sigma) = 0.
    \label{eq:ThirdLaw}
\end{align}

Information theoretically, the ground state becomes indistinguishable from a thermodynamic state at zero temperature. In contrast to the usual formulation based on entropy instead of relative entropy, the case of a degenerate ground state does not lead to a constant value on the right hand side of \eqref{eq:ThirdLaw}.
Let us first comment on \eqref{eq:ThirdLaw} for the case where $\sigma=\sigma_\text{c}$ is a canonical density matrix so that we can use eq.\ \eqref{eq:Relative_Entropy_Canonical}. Obviously, one needs to assume that $N(\rho)\equiv N(\sigma_\text{c})$, here. Moreover, $E(\rho_0)-E(\sigma_\text{c})$ must vanish for $\beta\to\infty$ faster than $1/\beta$, so that at zero temperature $S(\rho_0)=S(\sigma_\text{c})$. When $\sigma=\sigma_\text{gc}$ is a grand-canonical density matrix one can use \eqref{eq:Relative_Entropy_GrandCanonical} and one also needs $N(\rho_0)-N(\sigma_\text{gc})$ to vanish faster than $1/(\beta\mu)$ for $\beta\to\infty$ so that $S(\rho_0)=S(\sigma_\text{gc})$ at $T=0$. In both the canonical and the grand-canonical case, the formulation \eqref{eq:ThirdLaw} is then equivalent to Planck's formulation.

For completeness, let us now provide direct arguments for the validity of \eqref{eq:ThirdLaw}, concentrating for simplicity on the case of non-degenerate energy eigenstates and focusing on a canonical density matrix $\sigma_\text{c}$. In the energy eigenbasis the latter reads
\begin{align}
(\sigma_\text{c})_{n m} = \braket{n | \sigma_{\text{c}} | m} = \frac{1}{Z} e^{- \beta E_n} \delta_{n m} = q_n \, \delta_{n m},
\end{align}
where $q_n$ is the Gibbs distribution. The ground state density matrix $\rho_0$ is
\begin{align}
(\rho_0)_{n m} = \delta_{n 0} \, \delta_{m 0}.
\end{align}
Then the relative entropy becomes
\begin{align}
S(\rho_0 \| \sigma_{\text{c}})
&= \sum_n \left( \delta_{n 0} \ln \delta_{n 0} \right) - \sum_m \left(\delta_{m 0} \ln q_m  \right) \notag \\
&= - \ln q_0,
\end{align}
which is actually a general result for the relative entropy of the ground state to a density matrix that is diagonal in the energy eigenbasis. Now one can write
\begin{align}
\underset{T \rightarrow 0}{\lim} q_0 &= \underset{\beta \rightarrow \infty}{\lim} \frac{e^{- \beta E_0}}{\sum_n e^{- \beta E_n}} = 1 - \underset{\beta \rightarrow \infty}{\lim} \frac{\sum_{n>0} e^{- \beta E_n}}{\sum_n e^{- \beta E_n}} \notag \\
&= 1 - \underset{\beta \rightarrow \infty}{\lim} \frac{\sum_{n>0} e^{- \beta (E_n - E_0)}}{1 + \sum_{n>0} e^{- \beta (E_n - E_0)}}.
\end{align}
Note that $E_n - E_0 >0$ for all $n>0$. Indeed this leads to a confirmation of eq.\ \eqref{eq:ThirdLaw}.

\section{Conclusion and Outlook}
\label{chap:Conclusions}
In summary, we have investigated here to which extent the relation between thermodynamics and microscopic statistical physics can conceptually be formulated on the basis of {\it relative} entropy instead of entropy and the answer is in the affirmative. As a replacement for the principle of maximum entropy, that is usually taken as the conceptual starting point for the development of equilibrium thermodynamics, we have formulated a similar but new principle of minimum expected relative entropy. It is based on the construction of a (functional) integral measure on the space of probability distributions or density matrices in the classical and quantum formalism, respectively. This measure is actually Jeffreys prior, based on the square root of the determinant of the Fisher information metric (in the classical case) or of the quantum Fisher metric (in the quantum case), respectively. 

The microcanonical equilibrium state is then characterized as the from which all other states allowed by the constraints are least distinguishable -- measured in terms of relative entropy and Jeffreys prior. Based essentially on symmetry properties of the integral measure, we could show that this new variational principle leads to the standard microcanonical ensemble in a classical as well as in a quantum description.

From the microcanonical ensemble we have then re-developed also the canonical and grand-canonical ensemble, using relative entropy instead of entropy. This also includes alternative definitions of temperature and chemical potentials through specific derivatives of relative entropies.

An interesting point concerns also thermal fluctuations. While the traditional description going back to Einstein's work on critical opalescence is based on entropy, we have shown that an alternative formulation based on a relative entropy between a state with a given value of a fluctuating parameter and the thermal state, is possible and leads to an equivalent description. This alternative formulation has the advantage that it can directly be used in situations where entropy is infinite, but relative entropy is finite. 

Finally, we have also shown that the third law of thermodynamics can be formulated in terms of relative entropy instead of entropy itself.

Taken together, our results open on the one hand side a new perspective on foundational aspects of thermodynamics, for example our analysis suggests that it could be beneficial to think more often in terms of \textit{distinguishability} instead of \textit{missing information}. On the other side, the results also pave the ground for an application of thermodynamic concepts in situations where (entanglement) entropy is not finite, but relative (entanglement) entropy is. What we have specifically in mind here is the local description of a quantum field theory and its entanglement properties. The von Neumann entropy for the reduced density matrix of a spatial subregion (known as entanglement entropy) has severe ultraviolet divergences, while relative entanglement entropy is usually finite. With the formulation of thermodynamics in terms of relative entropy presented here, it might become possible to develop a better understanding of quantum field theories in close-to but out-of-equilibrium situations (see also ref.\ \cite{Floerchinger:2016gtl}) and to connect more directly to local thermal equilibrium approximations as they arise in the context of (relativistic) fluid dynamics. 

Other interesting questions for future work concern the operational definition of relative entropy directly within a purely thermodynamic context. In particular one may ask what is needed to define relative entropy as thermodynamic quantity and how this differs from the operational definition of entropy developed in refs.\ \cite{Lieb1999, Lieb2013, Lieb2014, Zanchini1986, Zanchini1988, Zanchini1992, Zanchini2008, Zanchini2010, Zanchini2011, Zanchini2014, Zanchini2019}.

\section*{Acknowledgements}
T.H. acknowledges useful discussions with Neil Dowling, Lars Heyen and Shahryar Ghaed Sharaf. This work is supported by the Deutsche Forschungsgemeinschaft (DFG, German Research Foundation) under Germany's Excellence Strategy EXC 2181/1 - 390900948 (the Heidelberg STRUCTURES Excellence Cluster), SFB 1225 (ISOQUANT) as well as FL 736/3-1.

%%%%%%%% Bibliography %%%%%%%%

%\bibliographystyle{unsrt} 
%\bibliographystyle{apsrev4-1} 
\bibliography{references.bib}

\end{document}